\documentclass[prb,twocolumn,superscriptaddress,floatfix,showpacs]{revtex4}
\usepackage{graphicx}           % eps figures
\begin{document}

\title{Room temperature ballistic transport in narrow graphene strips}

\author{D.\,Gunlycke}
\author{H.\,M.\,Lawler}
\author{C.\,T.\,White}
\affiliation{Naval Research Laboratory, Washington, DC 20375, USA}

\date{\today}

\begin{abstract}
We investigate electron-phonon couplings, scattering rates, and mean free paths in zigzag-edge graphene strips with widths of the order of $10$ nm.  Our calculations for these graphene nanostrips (GNSs) show both the expected similarity with single-wall carbon nanotubes (SWNTs) and suppression of the electron-phonon scattering due to a Dirichlet boundary condition that prohibits one major back\-scattering channel present in SWNTs.  Low-energy acoustic phonon scattering is exponentially small at room temperature due to the large phonon wavevector required for back\-scattering.  We find within our model that the electron-phonon mean free path is proportional to the width of the nano\-strip and is approximately $70$\,$\mu$m for an $11$-nm-wide nano\-strip.
\end{abstract}


\maketitle

\section{Introduction}
\label{sec:1}

Low-dimensional systems have long attracted attention in mesoscopic physics.  One area of such interest is one-dimensional conductors.  These quantum wires do not obey the usual macroscopic Ohm's Law.  Instead they exhibit ballistic or coherent diffusive electron transport.  In the former transport regime, the conductance is independent of the length of the quantum wire and is quantized in units of $2e^2/h$.  The requirement for ballistic transport is that the length of the quantum wire is shorter than its characteristic mean free path.  In metallic SWNTs at room temperature, low-energy mean free paths of order one micrometer have been observed \cite{Mann03,Park04}.  These observations are also confirmed by theory, which assumes that electron-phonon scattering is the dominant source of scattering \cite{Mann03,Park04} as the effects of static disorder are strongly suppressed \cite{Whit98,Ando98_1,Ando98_2,McEu99}.

In this paper, we present electron-phonon calculations of graphene nano\-strips terminated by hydrogen atoms.  The interior of the nano\-strips consists of an $sp^2$-bonded graphene honeycomb lattice, and therefore there is hope that the nano\-strips will inherit some of the special properties of graphene.  There are several studies of the electron properties of GNSs \cite{Fuji96,Naka96,Fuji97,Waka99,Miya99,Ramp99,Kusa03,Hiki03,AYama04,Kane05,Ezaw06,Pere06,Obra06,Brey06}, and recent experimental progress \cite{Berg06} in fabricating the materials will likely add further interest.  The nano\-strips can be made from a sheet of graphene using lithography.  Until recently, it was uncertain whether graphene would be thermodynamically stable or would spontaneously curl up into scrolls \cite{Novo05_1,Novo04}. Initial electron transport measurements on graphene have demonstrated high room temperature mobilities, suggesting that a sheet of graphene is relatively inert to interface scattering at the substrate surface \cite{Novo04,Novo05_2,Zhan05}.  From experiments, the graphene mean free path has been estimated to be $400$\,nm \cite{Novo04} and $600$\,nm \cite{Berg06}.  The mean free path has not shown significant temperature dependence, and therefore we would expect the electron-phonon mean free path to be even longer.

Unlike in graphene, the carriers in GNSs are confined to one dimension.  The electron energy dispersion is quantized in the transverse direction like the dispersion in carbon nanotubes.  However, there are two important differences between carbon nanotubes and nano\-strips.  First, metallic carbon nanotubes exhibit two channels (excluding the spin degree of freedom) at the Fermi level due to a periodic boundary condition in the transverse direction.  Nano\-strips, on the other hand, have a Dirichlet boundary condition and only a single channel.  Second, zigzag-edge nano\-strips have edge states which do not exist in carbon nanotubes \cite{Fuji96,Naka96}.  These edge states have energies close to the Fermi level, $\varepsilon_F$, and can therefore significantly affect low-bias electron transport.  On the other hand, the edge states are highly localized on the edges \cite{Fuji96,Naka96,Waka99} and have only small overlaps with the extended states, thus leaving the extended states in the dispersion relatively intact.  Similar edge localization has been seen in phonon energy dispersions \cite{TYama04}.  An important consequence of the presence of edge states is that they replace low-energy channels which would otherwise cause back\-scattering.

The next section describes the electron-phonon cross-sections in our model.  These results are then used in Sec.\,\ref{sec:3} where Fermi's Golden Rule is applied to calculate scattering rates and the electron-phonon mean free path.  Our findings are finally discussed in the concluding section.

\section{Electron-phonon couplings}
\label{sec:2}

\begin{figure}
  \includegraphics[angle=-90]{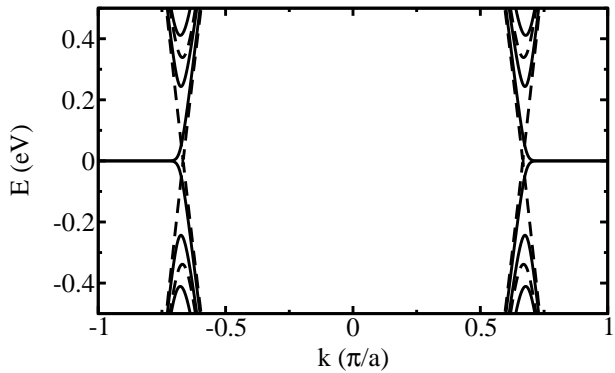}
  \caption{Electron energy dispersion of a graphene nano\-strip (solid bands).  This particular nano\-strip has a width of $N_{zz}=50$ interior zigzag lines which corresponds to approximately $11$\,nm.  The dispersion is compared to that of a $(25,25)$ carbon nanotube (dashed bands).  Due to the edge states there is only one channel close to the Fermi level which is located at $E=0$.}
  \label{f.1}
\end{figure}

We model the electron using the usual tight-binding Hamiltonian
\begin{equation}
  \label{e.1}
  H_\mathrm{el} = \sum_{\langle i,j\rangle}\gamma_{ij}~(c^{\dagger}_ic_j+\rm{h.\,c.}),
\end{equation}
where the summation is over all nearest neighbors $\langle i,j\rangle$, $\gamma_{ij}$ are hopping parameters, and $c^{\dagger}_i$ and $c_i$ are the fermionic creation and annihilation operators, respectively, on site $i$.  When there is no lattice deformation, we use the graphene $\pi$-orbital hopping parameter $\gamma_{ij} = \gamma_0 \approx -2.7$\,eV \cite{Whit98}.  The energy dispersion for a nano\-strip with $N_{zz}=50$ interior zigzag lines along the strip axis is shown in Fig.\,\ref{f.1}.
Close to the Fermi level, the strip has only a single channel in each direction, and the channels are separated by a wavevector approximately $q=(4/3)(\pi/a)$, where $a\approx 0.246$\,nm is the lattice spacing.  As a consequence, long-wavelength acoustic phonons cannot cause back\-scattering in zigzag-edge nano\-strips.

\begin{figure}
  \includegraphics[width=3.0in]{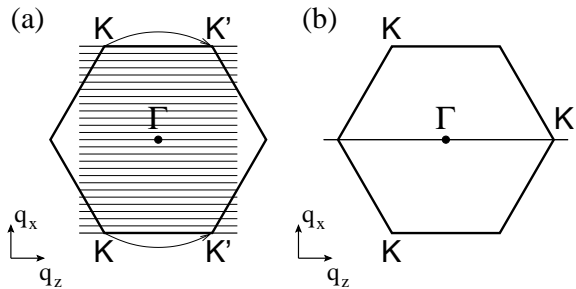}
  \caption{First Brillouin zone with (a) quantized electronic bands and (b) phonon branches satisfying $q_x=0$.  Phonons at the K-point can scatter an electron in state K into K'.}
  \label{f.2}
\end{figure}

\begin{figure}
  \includegraphics[width=3.0in]{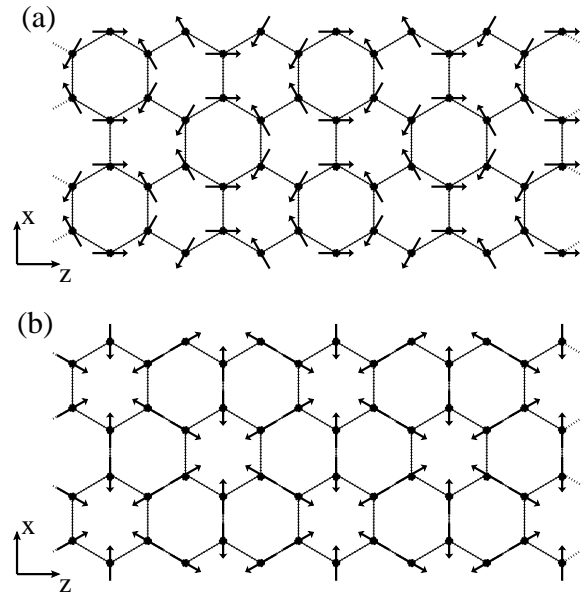}
  \caption{Lattice vibrations which cause non-negligible intraband electron-phonon coupling.  The two modes represent (a) a LA phonon and (b) a TO phonon at a K-point.  Note that due to the LA and LO degeneracy, the LA mode has been obtained in the limit $q\rightarrow (4/3)(\pi/a)$.}
  \label{f.3}
\end{figure}

In this paper, we restrict ourselves to single-phonon events with first-order variations in the hopping parameters.  We limit the calculation to the energy gap $\Delta E\approx (25/N_{zz})$\,eV in which there is only a single channel in each direction.  In graphene with a transverse boundary condition, there are only certain allowed transverse wavevectors as indicated in Fig.\,\ref{f.2}.
Only phonons with the same set of wavevectors can efficiently scatter electrons due to conservation of crystal momentum.  Furthermore, low-energy intraband electron back\-scattering can only be caused by phonons in the proximity of the K-point.  We will only consider the case where $q_x=0$ as all other equivalent K-points can be translated onto this band by a reciprocal lattice vector.  Because of the edge states in zigzag-edge nano\-strips, the quantization of the two-dimensional Brillouin zone is no longer exact although it still remains a good approximation in wide nano\-strips for $|k| < (2/3)(\pi/a)$, as the edge states are primarily located outside this wavevector window \cite{Fuji96,Naka96}.  There are six phonon branches with $q_x=0$.  In two of these branches, the phonon modes are entirely out-of-plane. These modes do not affect the bond length to first order and can therefore be neglected.  The remaining four phonon modes are longitudinal acoustic (LA), longitudinal optical (LO), transverse acoustic (TA), and transverse optical (TO) phonons.  The LA and TO phonons cause intraband scattering and the LO and TA phonons cause interband scattering between the conduction and valence bands.  As we will see below, the interband scattering can be minimized by doping.  This leaves the LA and TO branches with corresponding modes at a K-point shown in Fig.\,\ref{f.3}.

\begin{figure}
  \includegraphics{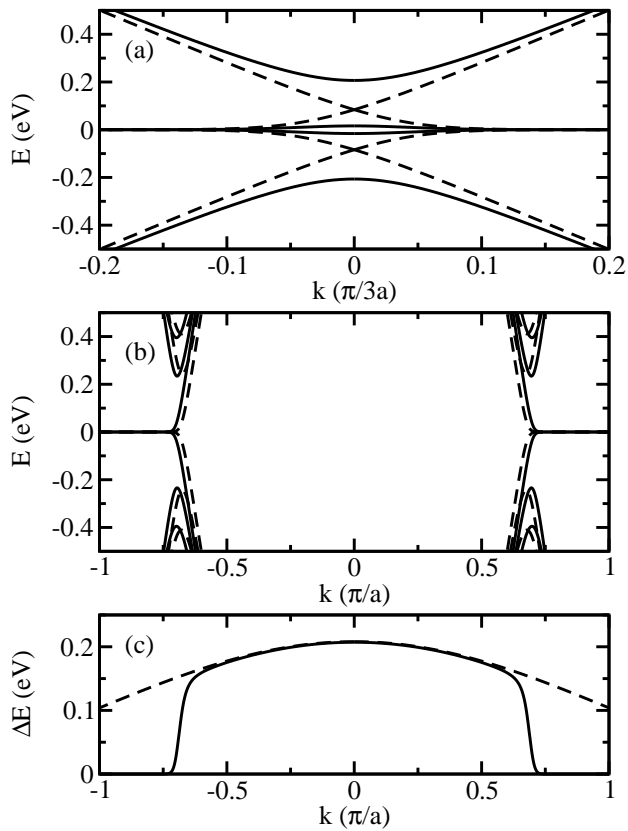}
  \caption{Effects due to a frozen TO phonon in a GNS ($N_{zz}=50$) lattice.  (a) The electron dispersion of the nano\-strip is calculated with the frozen phonon at the K-point (solid bands) and is compared to that with no frozen phonon (dashed bands).  Due to the frozen phonon, there are gaps at $k=0$ which reflect an electron-phonon coupling between $k=-(2/3)(\pi/a)$ and $k'=(2/3)(\pi/a)$.  (b) The same calculation with the frozen phonon at the $\Gamma$-point.  This phonon has a coupling which causes forward scattering.  (c) The energy difference between the electron dispersions with and without the frozen phonon at the $\Gamma$-point.  The energy difference in the nano\-strip (solid curve) is compared to that of a (25,25) armchair nanotube (dashed curve).  At low electron wavevectors the energy differences are similar, while at high wavevectors the energy difference in the nano\-strip is suppressed by its edge states.}
  \label{f.4}
\end{figure}

We can estimate the electron-phonon coupling in the nano\-strips by introducing a frozen phonon into the lattice.  The frozen phonon perturbs the lattice and causes local variations in the bond lengths, which in turn affect the nearest-neighbor hopping parameters.  We model this effect by expanding the Goodwin-Skinner-Pettifor scaling function for carbon \cite{Good91} and obtain hopping parameters of the form
\begin{equation}
  \label{e.2}
  \gamma_{ij} = \gamma_0 - \alpha~\hat{d}_{ij}\cdot(\vec{u}_i-\vec{u}_j),
\end{equation}
where the coupling constant is $\alpha\approx 52$\,eV/nm, $\hat{d}_{ij}$ is the unit vector in the direction to the nearest neighbor, and $\vec{u}_i-\vec{u}_j$ is the relative lattice displacement.  An LA or TO phonon at the K-point opens up a gap at $k=\pm (2/3)(\pi/a)$.  This effect can be seen in the reduced zone dispersion in Fig.\,\ref{f.4}(a), where $k=\pm (2/3)(\pi/a)$ has been folded back to $k=0$.
The gap is proportional to the electron-phonon coupling responsible for back\-scattering.  Although it does not reduce the electron-phonon mean free path, it is also worth looking at the TO phonon at the $\Gamma$-point.  This phonon, which only causes forward scattering, introduces shifts in the dispersion as illustrated in Fig.\,\ref{f.4}(b).  Fig.\,\ref{f.4}(c) shows the energy difference between the conduction bands calculated with and without the frozen TO phonon. The comparison to an armchair nanotube shows that the edge states are essentially unaffected by the introduction of a frozen phonon in the lattice.  This result is expected since a wavefunction of an edge state is dramatically different from that on an extended state.  The figure also demonstrates that the conduction band is similar to that in a nanotube for $|k| < (2/3)(\pi/a)$ which is also expected as the eigenstates become identical in the limit $N_{zz}\rightarrow\infty$.  The transition between the edge and extended states at $|k|\approx (2/3)(\pi/a)$ also become sharper for wider nano\-strips.  Motivated by Fig.\,\ref{f.4}(c) we take the electron-phonon matrix elements in graphene as an upper bound of those in GNSs and will from here on use the former, which we can estimate analytically.

The electron-phonon interaction in Fourier space can be expressed as
\begin{equation}
  \label{e.3}
  H_\mathrm{el-ph} = \sum_{\mu kq}M^\mu_{kq}c^\dagger_{k+q}c_k\phi^\mu_q,
\end{equation}
where $M^\mu_{kq}$ are the coupling matrix elements and $\phi^\mu_q$ is the phonon operator of mode $\mu$.  The latter is given by
\begin{equation}
  \label{e.4}
  \phi^\mu_q = \left(\frac{1}{2}\hbar\omega^\mu_q\right)^{1/2}(a^\mu_q+a^{\mu\dagger}_{-q}),
\end{equation}
where $a^\mu_q$, $a^{\mu\dagger}_{-q}$ are bosonic creation and annihilation operators, and $\omega^\mu_q$ are phonon frequencies.  The phonon dispersion is calculated numerically using a force constant model \cite{Dres00} with new parameters \cite{Grun02}.  Following the derivations of matrix elements in armchair carbon nanotubes \cite{Jish93,Maha03} we find the intraband matrix elements appropriate for nano\-strips
\begin{eqnarray}
  \label{e.5}
  M^\mathrm{LA}_{kq} &=& \frac{\mp i\sqrt{3}\alpha\sin{\frac{qa}{4}}\cos{\frac{(2k+q)a}{4}}}{\sqrt{2N_{zz}(L/a)m_\mathrm{c}}~\omega^\mathrm{LA}_q}(u_{AL}+u_{BL}),\nonumber\\
  M^\mathrm{TO}_{kq} &=& \frac{\pm \alpha\left[1+\cos{\frac{qa}{4}}\cos{\frac{(2k+q)a}{4}}\right]}{\sqrt{2N_{zz}(L/a)m_\mathrm{c}}~\omega^\mathrm{TO}_q}(u_{AT}-u_{BT}),\quad
\end{eqnarray}
and the interband matrix elements
\begin{eqnarray}
  \label{e.6}
  M^\mathrm{LO}_{kq} &=& \frac{\mp i\sqrt{3}\alpha\cos{\frac{qa}{4}}\sin{\frac{(2k+q)a}{4}}}{\sqrt{2N_{zz}(L/a)m_\mathrm{c}}~\omega^\mathrm{LO}_q}(u_{AL}-u_{BL}),\nonumber\\
  M^\mathrm{TA}_{kq} &=& \frac{\mp \alpha\sin{\frac{qa}{4}}\sin{\frac{(2k+q)a}{4}}}{\sqrt{2N_{zz}(L/a)m_\mathrm{c}}~\omega^\mathrm{TA}_q}(u_{AT}+u_{BT}),
\end{eqnarray}
where $L$ is the length of the strip, $m_\mathrm{c}$ is the mass of a carbon atom, and $(u_{AL},u_{AT},u_{BL},u_{BT})^\mathrm{T}$ is the polarization vector normalized to $2$.  The overall sign depends on whether $|k|$ is smaller or greater than $(2/3)(\pi/a)$ but is irrelevant for our scattering calculations.

\section{Scattering rates and mean free path}
\label{sec:3}

The electron-phonon scattering rate for an electron with wavevector $k$ can be calculated using the Fermi Golden Rule
\begin{equation}
  \label{e.7}
  \Gamma = \Gamma^\mathrm{(ab)} + \Gamma^\mathrm{(em)}
  = \sum_{\mu q} \left[\Gamma^\mathrm{(ab)}_{\mu kq}+\Gamma^\mathrm{(em)}_{\mu kq}\right],
\end{equation}
where the absorption and emission terms are
\begin{eqnarray}
  \label{e.8}
  \Gamma^\mathrm{(ab)}_{\mu kq} &=& \int\frac{2\pi}{\hbar}|\langle k+q,n^\mu_q-1|H_{el-ph}|k,n^\mu_q\rangle |^2 \nonumber \\
    &&~\times~\delta (E_{k+q}-E_k-\hbar\omega^\mu_q) \rho(E_{k+q})~\mathrm{d}E_{k+q} \nonumber \\
  &=& \pi|M^\mu_{kq}|^2\omega^\mu_qn^\mu_q~\rho(E_k+\hbar\omega^\mu_q),
\end{eqnarray}
and
\begin{eqnarray}
  \label{e.9}
  \Gamma^\mathrm{(em)}_{\mu kq} &=& \int\frac{2\pi}{\hbar}|\langle k+q,n^\mu_{-q}+1|H_{el-ph}|k,n^\mu_{-q}\rangle |^2 \nonumber \\
    &&~\times~\delta (E_{k+q}-E_k+\hbar\omega^\mu_q) \rho(E_{k+q})~\mathrm{d}E_{k+q} \nonumber \\
  &=& \pi|M^\mu_{kq}|^2\omega^\mu_q(1+n^\mu_q)~\rho(E_k-\hbar\omega^\mu_q),
\end{eqnarray}
where the unoccupied back\-scattering density of states is
\begin{equation}
  \label{e.10}
  \rho(E) = \frac{L}{\pi}\left|\frac{dk}{dE}\right|\left[1-\frac{1}{e^{(E-\mu)/k_\mathrm{B}T}+1}\right],
\end{equation}
where $\mu$ is the chemical potential.  The phonon occupation is given by the Bose-Einstein distribution
\begin{equation}
  \label{e.11}
  n^\mu_q = \frac{1}{e^{\hbar\omega^\mu_q/k_\mathrm{B}T}-1}.
\end{equation}
From Eqs.\,(\ref{e.5})-(\ref{e.11}) we find that the scattering rate is inversely proportional to the width of the nano\-strip.

\begin{figure}
  \includegraphics{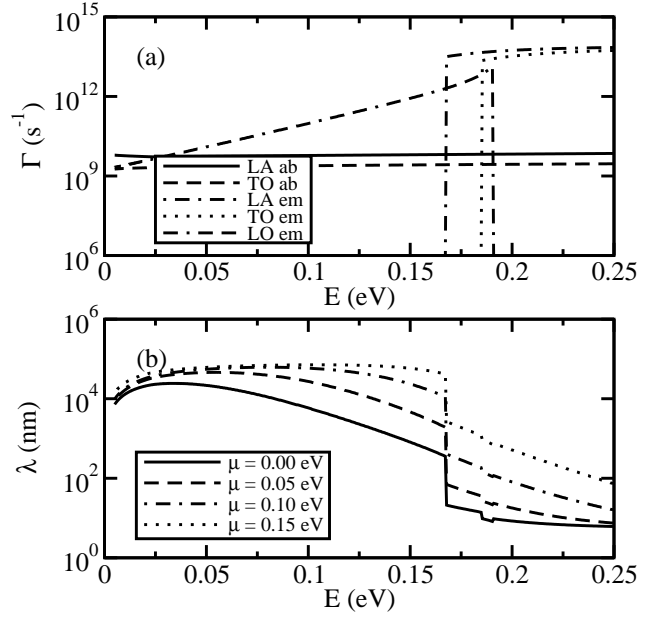}
  \caption{(a) Scattering rates of a $N_{zz}=50$ zigzag-edge GNS at $300$\,K as a function of electron energy.  The LA and TO phonons cause intraband and the LO phonon interband scattering.  Intraband phonon absorption occurs at all electron energies while intraband phonon emission requires energies larger than the threshold energy for the particular branch.  The threshold energies are about $0.16$\,eV and $0.18$\,eV for the LA and TO branches, respectively.  (b) The total mean free path at room temperature due to electron-phonon scattering.  The mean free path increases with the chemical potential since a higher chemical potential reduces the number of unoccupied electron states.}
  \label{f.5}
\end{figure}

Back\-scattering rates of an undoped ($\mu=0$) GNS at room temperature are shown in Fig.\,\ref{f.5}(a) as a function of initial electron energy.
Due to the large wavevector separation between the initial and final electron states the absorbed phonons must have considerable energies to ensure energy conservation.  These energies are similar to the threshold energies for LA and TO phonon emission.  The lowest threshold energy corresponds to a thermal energy of about $T_c =\hbar\omega^\mathrm{LA}_\mathrm{K}/k_\mathrm{B} \approx 1800$\,K. Therefore, phonon absorption is in a low-temperature regime ($T\ll T_c$) where the phonon occupation is exponentially small; $n^\mu_q\approx\exp(-\hbar\omega^\mu_q/k_\mathrm{B}T)$.  Because the absorption scattering rates scale with phonon occupation [see Eq.\,(\ref{e.8})] the back\-scattering due to phonon absorption is expected to be small, even at temperatures well above room temperature.  This effect does not occur in carbon nanotubes where low-energy long-wavelength acoustic phonons cause back\-scattering.  Back\-scattering due to phonon emission, on the other hand, is of the same order as in carbon nanotubes.

The electron-phonon mean free path is estimated by $\lambda = |v_i|/\Gamma$ where $v_i$ is the initial electron group velocity.  The mean free path at room temperature is shown in Fig.\,\ref{f.5}(b) for a few different chemical potentials.  At very small energies, the electrons move slowly which effectively shortens the mean free path.  As the electron energy increases, the electron speed approaches the Fermi velocity in graphene, $v_F$.  Together with increasing interband LO emission scattering, the mean free path forms a maximum.  This interband scattering can be reduced by doping the material to reduce the number of available states in the opposite band.  For $\mu = 0.15$\,eV, the interband scattering becomes small compared to intraband scattering for all energies.  The maximum mean free path can now be estimated by phonon absorption alone;
\begin{equation}
  \label{e.11}
  \lambda^\mathrm{(ab)}(N_{zz},T) = \lambda^\mathrm{(ab)}_0\frac{N_{zz}}{N^{(0)}_{zz}}~e^{T_c\left(\frac{1}{T}-\frac{1}{T_0}\right)},
\end{equation}
where $\lambda^\mathrm{(ab)}_0 \approx 70$\,$\mu$m, $N^{(0)}_{zz} = 50$, and $T_0 = 300$\,K.  Above the threshold for phonon emission the mean free path $\lambda^\mathrm{(em)}$ is of the order $10$\,nm which is comparable with that in carbon nanotubes\cite{Jave04}.

\section{Conclusions}
\label{sec:4}

Using the graphene sheet model [Eq.\,(\ref{e.1})], we have found within our model that electron-phonon scattering rates in zigzag-edge graphene nano\-stips is extraordinarily small for kinetic energies smaller than the phonon emission thresholds.  The reason for the small scattering rates is that long-wavelength acoustic phonons cannot cause back\-scattering of extended states in the lowest band.  Coupling to the edge states is expected to be small since overlaps between extended and edge wavefunctions are small and the phonons would have to have a highly exotic symmetry.  We have also shown that the electron-phonon mean free path is proportional to the width of the nano\-strip as long as excited bands cannot energetically be reached.  Wider nano\-strips, on the other hand, have the disadvantage that they reduce the single-channel window $\Delta E\approx (25/N_{zz})$\,eV, and for sufficently wide nano\-strips ($\Delta E\sim k_\mathrm{B}T$), electrons scattering to the next excited band will be appreciable, thereby reducing the mean free path.  The methodology presented in this paper can also be used to calculate electron-phonon scattering in armchair-edge nano\-strips.  The results are expected to be similar to zigzag nanotubes, although they might be irrelevant as armchair-edge nano\-strips are highly sensitive to edge disorder \cite{Ares06}.

Because the electron-phonon mean free path within the model is found to be extraordinary long in zigzag-edge nano\-strips, the actual electron mean free path is likely going to be limited either by disorder in the material, despite a recent prediction that zigzag-edge nano\-strips are relatively resistant to short-range, long-range, and edge disorders \cite{Ares06}, or by spin polarization effects \cite{Fuji96} at the edges which might open up an additional channel for backscattering.

%-------------------------------------------------------------------------%
\acknowledgments{DG and HML acknowledge support from the National Academies Research Associateship Programs.  This work was also supported by the Office of Naval Research both directly and through the Naval Research Laboratory.}

\end{document}